\documentclass[aps,twocolumn,groupedaddress,showpacs]{revtex4}
\usepackage[dvips]{graphicx}
\usepackage{amsmath}
\usepackage{amssymb}
\begin{document}
\bibliographystyle{prsty}

\date{\today}
\title{The Relic Problem of String Gas Cosmology}
\author{Diana Battefeld} 
\email{diana.battefeld(AT)helsinki.fi}
\affiliation{Helsinki Institute of Physics, P.O. Box 64, FIN-00014 Helsinki, Finland}
\affiliation{APC, UMR 7164, 10 rue Alice Domon et Leonie Duquet, 75205 Paris Cedex 13, France}
\author{Thorsten Battefeld}
\email{tbattefe(AT)princeton.edu}
\affiliation{Princeton University,
Department of Physics,
NJ 08544
}

\begin{abstract}
We discuss the relic problem of string gas cosmology (SGC) using gravitinos and magnetic monopoles as examples. Since SGC operates near or at the Hagedorn temperature, relics are produced copiously; in  the absence of dilution, their abundances are too large. A subsequent phase of reheating can solve the gravitino problem, but fails to dilute monopoles sufficiently. We propose a subsequent phase of inflation as the most natural solution to the monopole problem; unfortunately, inflation marginalizes almost all potential achievements of SGC, with the exception of a possible explanation of the dichotomy of space (why did only three dimensions inflate?).

\end{abstract}
\maketitle

\section{Introduction}

String or Brane Gas Cosmology (SGC) \cite{Kripfganz:1987rh,Brandenberger:1988aj,Tseytlin:1991xk} is an attempt to incorporate new, intrinsically stringy degrees of freedom into cosmology by means of a gas approximation while respecting T-duality \cite{Boehm:2002bm} (for reviews and comprehensive references see  \cite{Battefeld:2005av,Brandenberger:2008nx}) \footnote{The basic setup has been extended to incorporate branes, fluxes and more complicated topologies -- see \cite{Brandenberger:2008nx,Battefeld:2005av}.}. Several lines of research with different goals have emerged, among them are the stabilization of extra dimensions by massive, classical string gases  \cite{Battefeld:2004xw,Berndsen:2004tj,Berndsen:2005qq,Battefeld:2005av} or by quantum moduli trapping \cite{Watson:2003gf,Patil:2004zp,Patil:2005fi,Watson:2004aq,Kofman:2004yc,Berndsen:2005qq,Cremonini:2006sx,Greene:2007sa,Danos:2008pv}, isotropization of large dimensions \cite{Watson:2002nx},
 attempts to explain the late time acceleration in the visible dimensions via oscillations
of the extra dimensions caused by competition of a stabilizing string gas and matter \cite{Ferrer:2005hr,Ferrer:2008fp},
dark matter candidates \cite{Gubser:2004uh,Gubser:2004du,Nusser:2004qu,Battefeld:2004xw,Battefeld:2005av}, attempts to explain the dimensionality of the observed large dimensions \cite{Kripfganz:1987rh,Brandenberger:1988aj,Easther:2002qk,Easther:2003dd,Easther:2004sd,Danos:2004jz,Greene:2009gp}, bouncing/cyclic cosmologies caused by a coupling of a string gas with a modified gravity action \cite{Biswas:2006bs,Greene:2008hf},
and an attempt to provide a non-inflationary explanation of the nearly scale invariant spectrum of scalar perturbations by means of thermal fluctuations during a quasi-static Hagedorn phase \cite{Nayeri:2005ck,Brandenberger:2006xi,Brandenberger:2006vv,Brandenberger:2006pr}, which is assumed to be directly connected to an expanding, radiation dominated, $3+1$-dimensional FRW universe.  For shortcomings of some of these proposals and critical comments see the review \cite{Battefeld:2005av} as well as \cite{Kaloper:2006xw,Kaloper:2007pw}.

All proposals operate at or near the Hagedorn temperature $T_H$, the maximal attainable temperature in string theory ($T_H\sim 10^{16}\,\mbox{GeV}$, close to the grand-unified-theory (GUT) scale). Since this temperature is large, one should expect an over-production of unwanted relics if no precaution is taken to dilute them after the SGC-phase. 

In this note, we concretize this expectation by focusing on magnetic monopoles \cite{Zeldovich:1978wj,Preskill:1979zi} and gravitinos \cite{Ellis:1984eq}. We find that incorporating reheating via the decay of a relatively long lived particle, such as the s-axion in F-theory \cite{Heckman:2008jy} or moduli in G2-MSSM models \cite{Acharya:2008bk}, can dilute thermal relics such as gravitinos without spoiling the achievements of SGC, but it is insufficient to address the monopole problem. 

An incorporation of inflation can dilute all relics sufficiently at the expense of marginalizing achievements of SGC: internal dimensions that had been held together by a string gas are destabilized, dark matter candidates are diluted, and a non-inflationary explanation of a nearly scale invariant spectrum of scalar fluctuations becomes superfluous if inflation lasts long enough. Furthermore, since string gases do not drive inflation by themselves, further ingredients are needed.

\section{Relic Problems}
\subsection{Magnetic Monopoles}
Super-heavy magnetic monopoles 
 are an inevitable consequence of GUT's that include electromagnetism \cite{Zeldovich:1978wj,Preskill:1979zi,Guth:1979bh} (see also the textbook treatment in \cite{Mukhanov:2005sc}). During GUT phase transitions at least one monopole per horizon is produced due to causality, resulting in an overabundance of order \cite{Preskill:1979zi,Mukhanov:2005sc} $\Omega_M=\rho_{M}/\rho_{crit}\sim 10^{13}$ today, if the universe cooled down from $T_{GUT}\sim 10^{15}\,\mbox{GeV}$. Here, $\rho_{crit}=3H_0^2/(8\pi G)$ and $H_0$ is the Hubble parameter. A famous solution to this problem is inflation \cite{Guth:1980zm} \footnote{Super-heavy monopoles can also be produced during reheating \cite{Collins:1984vv,Lindblom:1984fk}, but observational bounds can usually be satisfied. There remains a danger of non-thermal overproduction during preheating, since energy is transferred to a few Fourier-modes before thermalization takes place.}, but the monopole problem resurfaces in SGC because $T_H\sim T_{GUT}$.   
 
 Non-inflationary attempts to address the monopole problem have been made, such as the Langacker-Pi mechanism \cite{Langacker:1980kd} \footnote{In the Langacker-Pi proposal \cite{Langacker:1980kd}, the universe becomes essentially super-conducting after breaking of the GUT-symmetry group but before the final breading of $SU_3\times U_1$. During this interval, monopoles are connected by flux tubes with anti-monopoles. These tubes shrink so that monopole-antimonopole pairs find each other and annihilate efficiently (the apparent violation of causality bounds \cite{Weinberg:1983uq} can be reconciled \cite{Everett:1984yc,Copeland:1987ht,Kibble:1990rg}, see also \cite{Holman:1992xs}).} or a recent proposal invoking primordial black holes to eliminate magnetic monopoles \cite{Stojkovic:2004hz}, but none of these proposals appears to be particularly compelling. 
 
A Hagedorn phase in SGC, as envisioned in  \cite{Nayeri:2005ck,Brandenberger:2006xi,Brandenberger:2006vv,Brandenberger:2006pr}, has to be followed by some mechanism to dilute monopoles, or else the model is ruled out \footnote{Another possibility is the absence of monopole production for some reason, i.e.~because the Hagedorn temperature is much lower than expected or breaking of the GUT gauge group does not permit monopole solutions. It remains to be seen whether or not concrete implementations of SGC within string theory have these peculiar features.}.

\subsection{Gravitinos}
Supersymmetric theories predict the existence of the gravitino, a Fermion of spin $3/2$ and the supersymmetric partner of the graviton that mediates supergravity interactions \cite{Freedman:1976xh}. The gravitino mass, $m_{3/2}$, originates from spontaneous supersymmetry breaking and its value ranges commonly from GeV-TeV; since it is long lived ($\tau_{3/2}\sim 4\times 10^5\,\mbox{s}\times (m_{3/2}/1\,\mbox{Tev})^{-3}$  if the dominant decay channel consists of a photon and its superpartner \cite{Holtmann:1998gd}), it is a natural candidate for dark matter. Even in the absence of primordial gravitinos, i.e. because primordial relics had been diluted by inflation, they are thermally produced during the radiation-dominated epoch; thus, they are an example of a thermal relic. Their abundance can be estimated to (see \cite{Heckman:2008jy} for a good review and \cite{Ellis:1984eq,Moroi:1993mb,Bolz:2000fu} for a more detailed analysis)
\begin{eqnarray}
\Omega_{3/2}^{T}h^2 \sim 2.7\times 10^{3}\frac{T_{3/2}^{min}}{10^{10}\,\mbox{GeV}}\frac{10\,\mbox{MeV}}{m_{3/2}}\frac{m_{\tilde{g}}}{1\,\mbox{TeV}}\,,
\end{eqnarray}
where $\Omega_{3/2}=\rho_{3/2}/\rho_{crit}$, $m_{\tilde{g}}$ is the gluino mass (we use $m_{\tilde{g}}=1\,\mbox{TeV}$ in the following), $T_{3/2}^{min}=\min(T^{max},T_{3/2}^{f})$, $T^{max}$ is the maximal temperature in the radiation dominated epoch and 
\begin{eqnarray}
T_{3/2}^{f}\sim 2\times 10^{10}\, \mbox{GeV}\left(\frac{m_{3/2}}{10\,\mbox{MeV}}\right)^2\left(\frac{1\,\mbox{TeV}}{m_{\tilde{g}}}\right)^2
\end{eqnarray}
is the gravitino freeze-out temperature. Since the gravitino density is bounded from above to prevent overclosure of the universe
\begin{eqnarray}
\Omega_{3/2}h^2\leq 0.1\,
\end{eqnarray}
a constraint on $T^{max}$ results; if $T_{3/2}^{f}<T^{max}$, we get $\Omega_{3/2}h^2\sim m_{3/2}/(2\,\mbox{keV})$ so that a low $m_{3/2}\leq 200\, \mbox{eV}$ is required \cite{Pagels:1981ke,Holtmann:1998gd}.  Focusing on $T_{3/2}^{f}>T^{max}$ leads to the upper bound
\begin{eqnarray}
T^{max}\leq\frac{1}{2.6}10^6\,\mbox{GeV} \left(\frac{m_{3/2}}{10\,\mbox{MeV}}\right)\,.
\end{eqnarray}

Another bound stems from the possible photo-dissociation of light elements created during big-bang-nucleosynthesis by decay products of gravitinos (assuming that the gravitino is not the lightest superparticle and thus stable) \cite{Khlopov:1984pf,Ellis:1984eq,Kawasaki:1994af,Holtmann:1998gd,Kohri:2005wn,Kawasaki:2008qe} if $\tau_{3/2}>1\,\mbox{s}$, 
\begin{eqnarray}
T^{max}&\leq & (10^6\,\mbox{--}\,10^8)\,\mbox{GeV}
\end{eqnarray}
 for $300\,\mbox{GeV} \lesssim m_{3/2} \lesssim 30\,\mbox{TeV}$, see table 2 in \cite{Kawasaki:2008qe}.

We see that the presence of thermally produced relics such as gravitinos imposes stringent constraints on the allowed maximal temperature in the radiation dominated epoch. 
As a consequence, upper limits on the reheating temperature after inflation result \cite{Kawasaki:2008qe}; these are satisfied in the old theory of reheating, but can cause tension for efficient preheating scenarios that put the reheating temperature close to the inflationary energy scale.

However, the problem is more severe in SGC, where the universe is thought to emerge from a thermal state near the Hagedorn temperature $T_H\sim 10^{16}\,\mbox{GeV}$. Since $T_H\gg 10^{8}\,\mbox{GeV}$, the above bounds are strongly violated if the universe simply cools down after leaving the Hagedorn phase, even if there were no primordial relics; thus, SGC requires a separate solution to the gravitino problem.

\section{Alleviating the Relic Problems?}
Two possibilities to evade relic problems are well known: a phase of reheating with $T_{RH}<T_{max}$ or a phase of inflation; both can dilute relics considerably. 

\subsection{Incorporating Reheating}
The decay of a relatively long lived particle species can reheat the universe and dilute relics by a factor of up to $10^{-5}$: to see this, consider radiation $\rho_r\propto a^{-4}$ (including relics) and a scalar field that oscillates in a quadratic potential so that 
$\rho_{\varphi}\propto a^{-3}$. Consider further that $\rho_\varphi=\rho_r\sim 10^{15}\,\mbox{GeV}$ after the SGC/Hagedorn-phase. If $\varphi$ decays around $\rho_\varphi\sim 1\, \mbox{GeV}$, the universe is reheats sufficiently for nucleosynthesis to commence, but to a low enough temperature to avoid any further production of unwanted relics. Since $\rho_{\varphi}/\rho_r\sim 10^5$ just before reheating, primordial relics and any other previously present particles are diluted considerably. Concrete implementations of this mechanism are the decay of the s-axion in F-theory \cite{Heckman:2008jy} or the decay of moduli in G2-MSSM models arising from M-theory compactifications \cite{Acharya:2008bk}.

The possible dilution of order $\sim 10^{-5}$ is sufficient to solve the gravitino problem \cite{Heckman:2008jy,Acharya:2008bk}. Since the late decaying field is conceivably produced towards the end of the Hagedorn phase, we expect that it carries fluctuations generated during this phase \footnote{The claim of a generated nearly scale invariant spectrum of scalar fluctuations during the Hagedorn phase \cite{Nayeri:2005ck,Brandenberger:2006vv,Brandenberger:2006pr} has to be taken with caution: the presence of a dynamical field, such as the dilation (which is expected to be present and dynamical \cite{Kaloper:2007pw}), results in a strongly scale dependent spectrum \cite{Kaloper:2006xw} (see also \cite{Brandenberger:2006pr}). In \cite{Biswas:2006bs} an attempt was made to use a higher derivative action combined with a string gas to give a realization of the structure formation mechanism of  \cite{Nayeri:2005ck,Brandenberger:2006vv,Brandenberger:2006pr} within a bouncing universe that avoids the criticisms of \cite{Kaloper:2006xw,Kaloper:2007pw}.}, a potentially desirable feature. Further, other achievements of SGC, such as moduli-stabilization or dark matter candidates, are not necessarily obliterated by this modest alteration of the universe's thermal history. However, the dilution caused by reheating is not enough to rid the universe of magnetic monopoles with an abundance of order $\Omega_M\sim 10^{13}$ -- they would still overclose the universe if not diluted by another method, such as the Langacker-Pi mechanism \cite{Langacker:1980kd}, or, more compelling, a phase of inflation. In the absence of such a mechanism, a Hagedorn phase in SGC is ruled out based on the monopole problem, even in the presence of reheating. 

\subsection{Incorporating Inflation}
A natural method to rid the universe of unwanted relics is inflation \cite{Guth:1980zm}, but it is challenging to incorporate inflation into SGC for a simple reason: a string gas does not cause inflation \cite{Tseytlin:1991xk}. If inflation is incorporated by other means after the Hagedorn phase, more than $N_{min}\equiv \ln(\Omega_M)/3\sim 12$ e-folds of accelerated expansion are required to sufficiently dilute the number density of magnetic monopoles. The string gas is necessarily subdominant during inflation and dilutes further, leading to the breakdown of the gas approximation; consequently, only a few horizon sized cosmic strings remain and almost all achievements of SGC vanish.  

Consider for instance the stabilization of extra dimensions by a string gas 
via quantum moduli trapping \cite{Watson:2003gf,Patil:2004zp,Patil:2005fi,Watson:2004aq,Kofman:2004yc,Greene:2007sa}. From a four dimensional point of view, size moduli appear as scalar fields; their stabilization requires a confining potential $V_{trap}$, which can originate from a string gas: for example, at the self dual radius certain states (a combination of winding and momentum modes) become light, get produced and can trap the radion \cite{Patil:2004zp,Patil:2005fi,Battefeld:2005av}. However, a subsequent phase of inflation removes this stabilization mechanism, as mentioned in \cite{Battefeld:2005av}, since the trapping potential is proportional to the number density of strings and redshifts like matter, $V_{trap}\propto a^{-3}$. Thus, this achievement of SGC quickly vanishes during inflation, and, in the above example, internal dimensions can grow if they are not stabilized by other means.

If inflation does not last much longer than $N_{min}$, one might still hope that the degree of freedom causing inflation inherited fluctuations on observable large scales during the Hagedorn phase, thus leaving a loophole for the proposal of \cite{Nayeri:2005ck,Brandenberger:2006vv,Brandenberger:2006pr} to operate. However, since the duration of inflation needs to be fine tuned \footnote{An extended phase of inflation is unusual if low scale inflation driven by fields within the MSSM is incorporated (see \cite{Allahverdi:2008zz} for a recent review). Thus, achieving only a few e-folds of inflation might be natural in some models and not fine tuned.}, and inflation naturally generates the needed nearly scale invariant spectrum of scalar fluctuations on large scales if it lasts $\sim 50$ e-folds or longer, one may justifiably use Ockham's razor to rid the scenario of the Hagedorn phase entirely.

Nevertheless, there is one line of research within SGC that remains interesting, even in the presence of inflation: to provide the initial conditions for inflation, that is, to explain why three dimensions inflated while the extra dimensions did not. Hence, it would be interesting to extend research primarily focused on explaining why only three dimensions became large in the presence of a string gas \cite{Kripfganz:1987rh,Brandenberger:1988aj,Easther:2002qk,Easther:2003dd,Easther:2004sd} to frameworks that include additional ingredients capable of driving inflation.  

One such framework was proposed in \cite{Shuhmaher:2005pw,Battefeld:2006cn} (see also \cite{Brandenberger:2003ge,Biswas:2005ap,Easson:2005ug} for related work on incorporating inflation by means of brane gases) where anisotropic inflation (as argued for in \cite{Battefeld:2005wv}) is incorporated: the multi-dimensional universe starts
out small and hot, with our three dimensions compactified
on a torus and the extra dimensions on an orbifold
of the same size. The pre-inflationary expansion 
is governed by topological defects (p-branes) in the bulk \cite{Shuhmaher:2005pw}. As the
universe expands isotropically due to the gas of p-branes,
the energy density stored in the gas is diluted until additional
weak forces come into play, changing the overall
dynamics. Branes pinned to the orbifold fixed
planes, which can exhibit an attractive force \footnote{The inter-brane potential is postulated in \cite{Shuhmaher:2005pw,Battefeld:2006cn}.}, eventually
cause a contraction of the extra dimensions while our
dimensions inflate. The inflaton is identified with the radion -- its large initial value is explained by 
the pre-inflationary bulk expansion. Inflation ends once the internal dimensions approach the self dual radius where moduli trapping \footnote{Even without a subsequent phase of inflation, the string gas may dilute too much to guarantee stabilization of internal dimensions at late times, see \cite{Battefeld:2006cn}.} and
(p)re-heating takes place. Temperatures remain below $T_H$, and an over-production of relics is avoided. However, the dichotomy between inflating and deflating dimensions is put in by hand, making a dynamical explanation desirable.

\section{Conclusions}
In this note, we briefly reviewed relic problems caused by magnetic monopoles and gravitinos, which pose a challenge for string gas cosmology (SGC). Since the universe is thought to emerge from a thermal state near the Hagedorn temperature in many proposals within the framework of SGC, we argue that the production of either relic, if not diluted by some means, rules out such a setup \footnote{The same line of reasoning rules out bouncing/cyclic models of the universe, if the bounce occurs at Planckian densities (as expected if the bounce is caused by quantum-gravity effects) and the bounce connects directly to an expanding, radiation dominated FRW universe.}. A late stage of reheating, albeit successful in providing a solution to the gravitino problem, does not dilute magnetic monopoles sufficiently. The latter ones are generic predictions of GUT phase transitions after a Hagedorn phase. Incorporating a phase of inflation is the most compelling solution to 
the monopole problem, but it comes at the price of marginalizing almost all achievements of SGC, even though finding a possible explanation for the dimensionality of spacetime (why did only three dimensions start to inflate?) remains a worthwhile endeavor. Non-inflationary solutions to the monopole problem, such as the Langacker-Pi mechanism, are less compelling, but might be invoked if one is adamant to avoid inflation. 

We did not address other shortcomings of SGC, such as the flatness or the entropy problem, which are also most readily solved by incorporating inflation \cite{Shuhmaher:2008xp} (see \cite{Brandenberger:2005qj,Shuhmaher:2008xp} for an attempt to address these problems without invoking inflation). We conclude that SGC is not yet an alternative to inflation, since it appears to require inflation, or a plethora of other mechanisms, to solve several well known problems of standard big-bang cosmology.

\begin{acknowledgments}
 We thank S.~Patil, P.~Steinhardt and S.~Watson for discussions and R.~Brandenberger, T.~Biswas, D.~Easson and A.~Frey for comments on the draft. T.~B.~is supported by the Council on Science and Technology at Princeton University and acknowledges hospitality at the APC (Paris, France) and the Perimeter Institute. D.~B.~is supported by the EU EP6 Marie Curie Research and Training Network `UniverseNet' (MRTN-CT-2006-035863) and acknowledges hospitality at Princeton University.
\end{acknowledgments}


\begin{thebibliography}{99}


\bibitem{Kripfganz:1987rh}
  J.~Kripfganz and H.~Perlt,
  Class.\ Quant.\ Grav.\  {\bf 5}, 453 (1988).
 
  
\bibitem{Brandenberger:1988aj}
  R.~H.~Brandenberger and C.~Vafa,
  Nucl.\ Phys.\  B {\bf 316}, 391 (1989).
  
\bibitem{Tseytlin:1991xk}
  A.~A.~Tseytlin and C.~Vafa,
  Nucl.\ Phys.\  B {\bf 372}, 443 (1992)
  [arXiv:hep-th/9109048].


\bibitem{Boehm:2002bm}
  T.~Boehm and R.~Brandenberger,
  JCAP {\bf 0306}, 008 (2003)
  [arXiv:hep-th/0208188].

\bibitem{Battefeld:2005av}
  T.~Battefeld and S.~Watson,
  Rev.\ Mod.\ Phys.\  {\bf 78}, 435 (2006)
  [arXiv:hep-th/0510022].

\bibitem{Brandenberger:2008nx}
  R.~H.~Brandenberger,
  arXiv:0808.0746 [hep-th].

\bibitem{Battefeld:2004xw}
  T.~Battefeld and S.~Watson,
  JCAP {\bf 0406}, 001 (2004)
  [arXiv:hep-th/0403075].

\bibitem{Berndsen:2004tj}
  A.~J.~Berndsen and J.~M.~Cline,
  Int.\ J.\ Mod.\ Phys.\  A {\bf 19}, 5311 (2004)
  [arXiv:hep-th/0408185].


\bibitem{Berndsen:2005qq}
  A.~Berndsen, T.~Biswas and J.~M.~Cline,
  JCAP {\bf 0508}, 012 (2005)
  [arXiv:hep-th/0505151].
  
\bibitem{Cremonini:2006sx}
  S.~Cremonini and S.~Watson,
  Phys.\ Rev.\  D {\bf 73}, 086007 (2006)
  [arXiv:hep-th/0601082].

\bibitem{Watson:2003gf}
  S.~Watson and R.~Brandenberger,
  JCAP {\bf 0311}, 008 (2003)
  [arXiv:hep-th/0307044].

\bibitem{Patil:2004zp}
  S.~P.~Patil and R.~Brandenberger,
  Phys.\ Rev.\  D {\bf 71}, 103522 (2005)
  [arXiv:hep-th/0401037].

\bibitem{Patil:2005fi}
  S.~P.~Patil and R.~H.~Brandenberger,
  JCAP {\bf 0601}, 005 (2006)
  [arXiv:hep-th/0502069].
  

\bibitem{Watson:2004aq}
  S.~Watson,
  Phys.\ Rev.\  D {\bf 70}, 066005 (2004)
  [arXiv:hep-th/0404177].
  
\bibitem{Kofman:2004yc}
  L.~Kofman, A.~Linde, X.~Liu, A.~Maloney, L.~McAllister and E.~Silverstein,
  JHEP {\bf 0405}, 030 (2004)
  [arXiv:hep-th/0403001].
  
\bibitem{Greene:2007sa}
  B.~Greene, S.~Judes, J.~Levin, S.~Watson and A.~Weltman,
  JHEP {\bf 0707}, 060 (2007)
  [arXiv:hep-th/0702220].

\bibitem{Danos:2008pv}
  R.~J.~Danos, A.~R.~Frey and R.~H.~Brandenberger,
  Phys.\ Rev.\  D {\bf 77}, 126009 (2008)
  [arXiv:0802.1557 [hep-th]].


\bibitem{Watson:2002nx}
  S.~Watson and R.~H.~Brandenberger,
  Phys.\ Rev.\  D {\bf 67}, 043510 (2003)
  [arXiv:hep-th/0207168].
  
\bibitem{Ferrer:2005hr}
  F.~Ferrer and S.~Rasanen,
  JHEP {\bf 0602}, 016 (2006)
  [arXiv:hep-th/0509225].

\bibitem{Ferrer:2008fp}
  F.~Ferrer, T.~Multamaki and S.~Rasanen,
  JHEP {\bf 0904}, 006 (2009)
  [arXiv:0812.4182 [hep-th]].
  
\bibitem{Gubser:2004uh}
  S.~S.~Gubser and P.~J.~E.~Peebles,
  Phys.\ Rev.\  D {\bf 70}, 123510 (2004)
  [arXiv:hep-th/0402225].
  

\bibitem{Gubser:2004du}
  S.~S.~Gubser and P.~J.~E.~Peebles,
  Phys.\ Rev.\  D {\bf 70}, 123511 (2004)
  [arXiv:hep-th/0407097].


\bibitem{Nusser:2004qu}
  A.~Nusser, S.~S.~Gubser and P.~J.~E.~Peebles,
  Phys.\ Rev.\  D {\bf 71}, 083505 (2005)
  [arXiv:astro-ph/0412586].


\bibitem{Easther:2002qk}
  R.~Easther, B.~R.~Greene, M.~G.~Jackson and D.~N.~Kabat,
  Phys.\ Rev.\  D {\bf 67}, 123501 (2003)
  [arXiv:hep-th/0211124].
  
\bibitem{Easther:2003dd}
  R.~Easther, B.~R.~Greene, M.~G.~Jackson and D.~N.~Kabat,
  JCAP {\bf 0401}, 006 (2004)
  [arXiv:hep-th/0307233].
  
\bibitem{Easther:2004sd}
  R.~Easther, B.~R.~Greene, M.~G.~Jackson and D.~N.~Kabat,
  JCAP {\bf 0502}, 009 (2005)
  [arXiv:hep-th/0409121].
  
  
\bibitem{Danos:2004jz}
  R.~Danos, A.~R.~Frey and A.~Mazumdar,
  Phys.\ Rev.\  D {\bf 70}, 106010 (2004)
  [arXiv:hep-th/0409162].
  
\bibitem{Greene:2009gp}
  B.~Greene, D.~Kabat and S.~Marnerides,
  arXiv:0908.0955 [hep-th].


\bibitem{Biswas:2006bs}
  T.~Biswas, R.~Brandenberger, A.~Mazumdar and W.~Siegel,
  JCAP {\bf 0712}, 011 (2007)
  [arXiv:hep-th/0610274].

\bibitem{Greene:2008hf}
  B.~Greene, D.~Kabat and S.~Marnerides,
  arXiv:0809.1704 [hep-th].

\bibitem{Nayeri:2005ck}
  A.~Nayeri, R.~H.~Brandenberger and C.~Vafa,
  Phys.\ Rev.\ Lett.\  {\bf 97}, 021302 (2006)
  [arXiv:hep-th/0511140].

\bibitem{Brandenberger:2006xi}
  R.~H.~Brandenberger, A.~Nayeri, S.~P.~Patil and C.~Vafa,
  Phys.\ Rev.\ Lett.\  {\bf 98}, 231302 (2007)
  [arXiv:hep-th/0604126].

\bibitem{Brandenberger:2006vv}
  R.~H.~Brandenberger, A.~Nayeri, S.~P.~Patil and C.~Vafa,
  Int.\ J.\ Mod.\ Phys.\  A {\bf 22}, 3621 (2007)
  [arXiv:hep-th/0608121].

\bibitem{Brandenberger:2006pr}
  R.~H.~Brandenberger {\it et al.},
  JCAP {\bf 0611}, 009 (2006)
  [arXiv:hep-th/0608186].
  


\bibitem{Kaloper:2006xw}
  N.~Kaloper, L.~Kofman, A.~Linde and V.~Mukhanov,
  JCAP {\bf 0610}, 006 (2006)
  [arXiv:hep-th/0608200].
  
\bibitem{Kaloper:2007pw}
  N.~Kaloper and S.~Watson,
  Phys.\ Rev.\  D {\bf 77}, 066002 (2008)
  [arXiv:0712.1820 [hep-th]].
  
 
\bibitem{Zeldovich:1978wj}
  Y.~B.~Zeldovich and M.~Y.~Khlopov,
  Phys.\ Lett.\  B {\bf 79}, 239 (1978).

\bibitem{Preskill:1979zi}
  J.~Preskill,
  Phys.\ Rev.\ Lett.\  {\bf 43}, 1365 (1979).


\bibitem{Ellis:1984eq}
  J.~R.~Ellis, J.~E.~Kim and D.~V.~Nanopoulos,
  Phys.\ Lett.\  B {\bf 145}, 181 (1984).

\bibitem{Heckman:2008jy}
  J.~J.~Heckman, A.~Tavanfar and C.~Vafa,
  arXiv:0812.3155 [hep-th].
  
\bibitem{Acharya:2008bk}
  B.~S.~Acharya, P.~Kumar, K.~Bobkov, G.~Kane, J.~Shao and S.~Watson,
  JHEP {\bf 0806}, 064 (2008)
  [arXiv:0804.0863 [hep-ph]].

\bibitem{Guth:1979bh}
  A.~H.~Guth and S.~H.~H.~Tye,
  Phys.\ Rev.\ Lett.\  {\bf 44}, 631 (1980)
  [Erratum-ibid.\  {\bf 44}, 963 (1980)].

\bibitem{Mukhanov:2005sc}
  V.~Mukhanov,
{\it  Cambridge, UK: Univ. Pr. (2005) 421 p}

 
\bibitem{Guth:1980zm}
  A.~H.~Guth,
  Phys.\ Rev.\  D {\bf 23}, 347 (1981).
  
\bibitem{Lindblom:1984fk}
  P.~R.~Lindblom and P.~J.~Steinhardt,
  Phys.\ Rev.\  D {\bf 31}, 2151 (1985).
  
\bibitem{Collins:1984vv}
  W.~Collins and M.~S.~Turner,
  Phys.\ Rev.\  D {\bf 29}, 2158 (1984).
    
\bibitem{Langacker:1980kd}
  P.~Langacker and S.~Y.~Pi,
  Phys.\ Rev.\ Lett.\  {\bf 45}, 1 (1980).


\bibitem{Weinberg:1983uq}
  E.~J.~Weinberg,
  Phys.\ Lett.\  B {\bf 126}, 441 (1983).

\bibitem{Everett:1984yc}
  A.~E.~Everett, T.~Vachaspati and A.~Vilenkin,
  Phys.\ Rev.\  D {\bf 31}, 1925 (1985).

\bibitem{Copeland:1987ht}
  E.~J.~Copeland, D.~Haws, T.~W.~B.~Kibble, D.~Mitchell and N.~Turok,
  Nucl.\ Phys.\  B {\bf 298}, 445 (1988).
  
\bibitem{Kibble:1990rg}
  T.~W.~B.~Kibble and E.~J.~Weinberg,
  Phys.\ Rev.\  D {\bf 43}, 3188 (1991).


\bibitem{Holman:1992xs}
  R.~Holman, T.~W.~B.~Kibble and S.~J.~B.~Rey,
  Phys.\ Rev.\ Lett.\  {\bf 69}, 241 (1992)
  [arXiv:hep-ph/9203209].

\bibitem{Stojkovic:2004hz}
  D.~Stojkovic and K.~Freese,
  Phys.\ Lett.\  B {\bf 606}, 251 (2005)
  [arXiv:hep-ph/0403248].
  

\bibitem{Freedman:1976xh}
  D.~Z.~Freedman, P.~van Nieuwenhuizen and S.~Ferrara,
  Phys.\ Rev.\  D {\bf 13}, 3214 (1976).
  
\bibitem{Holtmann:1998gd}
  E.~Holtmann, M.~Kawasaki, K.~Kohri and T.~Moroi,
  Phys.\ Rev.\  D {\bf 60}, 023506 (1999)
  [arXiv:hep-ph/9805405].


\bibitem{Moroi:1993mb}
  T.~Moroi, H.~Murayama and M.~Yamaguchi,
  Phys.\ Lett.\  B {\bf 303}, 289 (1993).

\bibitem{Bolz:2000fu}
  M.~Bolz, A.~Brandenburg and W.~Buchmuller,
  Nucl.\ Phys.\  B {\bf 606}, 518 (2001)
  [Erratum-ibid.\  B {\bf 790}, 336 (2008)]
  [arXiv:hep-ph/0012052].
  
\bibitem{Pagels:1981ke}
  H.~Pagels and J.~R.~Primack,
  Phys.\ Rev.\ Lett.\  {\bf 48}, 223 (1982).
  
  
 \bibitem{Khlopov:1984pf}
  M.~Y.~Khlopov and A.~D.~Linde,
  Phys.\ Lett.\  B {\bf 138} (1984) 265.


\bibitem{Kawasaki:1994af}
  M.~Kawasaki and T.~Moroi,
  Prog.\ Theor.\ Phys.\  {\bf 93}, 879 (1995)
  [arXiv:hep-ph/9403364].
  
\bibitem{Kohri:2005wn}
  K.~Kohri, T.~Moroi and A.~Yotsuyanagi,
  Phys.\ Rev.\  D {\bf 73}, 123511 (2006)
  [arXiv:hep-ph/0507245].
  
\bibitem{Kawasaki:2008qe}
  M.~Kawasaki, K.~Kohri, T.~Moroi and A.~Yotsuyanagi,
  Phys.\ Rev.\  D {\bf 78}, 065011 (2008)
  [arXiv:0804.3745 [hep-ph]].

\bibitem{Allahverdi:2008zz}
  R.~Allahverdi,
  Mod.\ Phys.\ Lett.\  A {\bf 23}, 2799 (2008)
  [arXiv:0812.3628 [hep-ph]].
   
\bibitem{Battefeld:2006cn}
  T.~Battefeld and N.~Shuhmaher,
  Phys.\ Rev.\  D {\bf 74}, 123501 (2006)
  [arXiv:hep-th/0607061].

  
\bibitem{Shuhmaher:2005pw}
  N.~Shuhmaher and R.~Brandenberger,
  Phys.\ Rev.\ Lett.\  {\bf 96}, 161301 (2006)
  [arXiv:hep-th/0512056].


\bibitem{Brandenberger:2003ge}
  R.~Brandenberger, D.~A.~Easson and A.~Mazumdar,
  Phys.\ Rev.\  D {\bf 69}, 083502 (2004)
  [arXiv:hep-th/0307043].


\bibitem{Biswas:2005ap}
  T.~Biswas, R.~Brandenberger, D.~A.~Easson and A.~Mazumdar,
  Phys.\ Rev.\  D {\bf 71}, 083514 (2005)
  [arXiv:hep-th/0501194].


\bibitem{Easson:2005ug}
  D.~A.~Easson and M.~Trodden,
  Phys.\ Rev.\  D {\bf 72}, 026002 (2005)
  [arXiv:hep-th/0505098].


\bibitem{Battefeld:2005wv}
  T.~J.~Battefeld, S.~P.~Patil and R.~H.~Brandenberger,
  Phys.\ Rev.\  D {\bf 73}, 086002 (2006)
  [arXiv:hep-th/0509043].

\bibitem{Brandenberger:2005qj}
  R.~Brandenberger and N.~Shuhmaher,
  JHEP {\bf 0601}, 074 (2006)
  [arXiv:hep-th/0511299].
  
\bibitem{Shuhmaher:2008xp}
  N.~Shuhmaher,
  arXiv:0803.0216 [hep-ph].
  




 




 
   
\end{thebibliography}
\end{document}